# Multichannel remote polarization control enabled by nanostructured Liquid Crystalline Networks

Simone Zanotto[1],[*], Fabrizio Sgrignuoli[2,3], Sara Nocentini[2,4,5], Daniele Martella[4,6,7], Camilla Parmeggiani[4,5,6,7], and Diederik S. Wiersma[2,4,5,6]

1. Istituto di Nanoscienze - CNR, Laboratorio NEST, Piazza San Silvestro 12, 56127 Pisa, Italy
2. Dipartimento di Fisica ed Astronomia, Università degli Studi di Firenze, Via Sansone 1, 50019 Sesto Fiorentino (FI), Italy
3. Department of Electrical and Computer Engineering, Boston University, 8 Saint Mary's Street, Boston, MA, 02215-1300, United States of America
4. LENS - European Laboratory for Nonlinear Spectroscopy, Università degli Studi di Firenze, Via Nello Carrara 1, 50019 Sesto Fiorentino (FI), Italy
5. Istituto Nazionale di Ricerca Metrologica, Strada delle Cacce 91, 10135 Torino, Italy
6. Istituto Nazionale di Ottica – CNR, Via Nello Carrara 1, 50019 Sesto Fiorentino (FI), Italy
7. Dipartimento di Chimica "Ugo Schiff", Università degli Studi di Firenze, Via della Lastruccia 3-13, 50019 Sesto Fiorentino (FI), Italy

**Abstract**

**In this article we demonstrate that a grating fabricated through nanoscale volumetric crosslinking of a liquid crystalline polymer enables remote polarization control over the diffracted channels. This functionality is a consequence of the responsivity of liquid crystal networks upon light stimuli. Tuning the photonic response of the device is obtained thanks to both a refractive index and a shape change of the grating elements induced by a molecular rearrangement under irradiation. In particular, the material anisotropy allows for nontrivial polarization state management over multiple beams. Absence of any liquid component and a time response down to 0.2 milliseconds make our device appealing in the fields of polarimetry and optical communications.**

Polarization is a key property of light, recognized since the times of Newton, who supported the idea that light has "sides" [1]. Intriguingly, it results that animals and plants started to make use of polarization well before, as certain polarization-dependent structural colors and specific receptors clearly testify [2,3]. Presently, man-made polarization handling devices find their use in several aspects of science and technology, such as polarimetric sensing [4], augmented reality devices [5], astrophysics [6], and optical communications [7,8].

In this regard, generation, detection and conversion of polarization states is an essential operation that, in most of the cases, relies on the birefringent properties of inorganic crystals or liquid crystals. Although these are mature technologies, the first suffers from bulkiness, and the second requires special care to confine the liquid phase. Moreover, their spectral response cannot be easily tuned beyond the intrinsic response dictated by the material. Photonic nanostructures solve this issue, as they allow to obtain optical properties on demand by appropriate subwavelength patterning of dielectric or metallic materials [9]. By these means it is possible to implement very general polarization response from isotropic materials [10–13], and to develop division-of-amplitude static polarimeters which show unprecedented speed and compactness [14,15]. Moreover, these operations can be reconfigured statically or dynamically by exploiting thermo-optic, electro-optic, electrochromic, or optomechanic effects [16–25]. While several proposed devices rely on plasmonic elements to exploit field enhancement effects, dielectric photonic structures are also of great interest because they are almost lossless. In this view, a new class of material, named Liquid Crystalline Networks (LCNs), has acquired a lot of attention within the photonic community due to its unique properties. In fact, these materials exhibit low optical loss, intrinsic optical anisotropy, machinability to subwavelength precision, and tunability. LCNs are polymeric networks where liquid crystalline chains are interconnected, giving rise to an elastomeric material [26,27] that sustains different phase transitions. The materials employed in the present work – i.e., polyacrylates – show a nematic-paranematic transition, across which birefringence, stress, and strain change, with a direct impact on the photonic properties of a LCN nanostructure [28,29]. While certain photonic devices based on nanostructured LCN have been already demonstrated [30–36], our aim is here to illustrate a different application: a remotely controllable multichannel polarization conversion element. It is indeed the important feature of remote, non-invasive tuning that makes our device a significant advancement with respect on previous findings on multichannel photonic components [37].

[*] email: simone.zanotto@nano.cnr.it

The device concept is schematized in Fig. 1a: an anisotropic diffraction grating is fabricated by local polymerization and crosslinking of a liquid crystal monomer mixture. The technique employed is that of two-photon polymerization, which enables LCN patterning with a resolution up to 160 nm [38]. Details of the employed material and of the fabrication process are illustrated in the Supplementary Material. Thanks to the stable nematic phase at room temperature, the photopolymerization process "freezes" the molecule orientation - irrespectively of the stripe writing direction – in the polymerized structures (Fig. 1b). In this work, homogeneous alignment of the nematic molecules has been adopted and the grating stripe orientation has been chosen at 45° with respect to the LC director. This choice results in an array of highly anisotropic refractive stripes on the glass surface. Therefore, light beams emerging from the grating (both transmitted, reflected and diffracted) have a polarization state with different tilt and ellipticity with respect to that of the incident

beam – even at normal incidence (Fig. 1a). Moreover, by appropriate tuning of the stripe geometrical parameters (i.e., width, thickness, periodicity), it is in principle possible to implement a general polarization state transformer that acts independently over several diffracted beams, i.e., over several channels. Noticeably, polymer photonic platforms allow for an easy incorporation of controlled gain/loss elements, which would enhance the device functionality thanks to non-Hermitian optics concepts [39,40].

The polarization behavior of our grating is dynamically tunable thanks to the responsivity of the LCN, whose shape and refractive index change as the nematic-paranematic phase transition is crossed. When this transition occurs, the molecular structure of the LCN becomes more isotropic, and the grating polarization response shifts towards the ordinary one. An important feature of LCNs is that their response can be remotely controlled by means of a light stimulus [41,42]. The mechanism behind the photo-response of the reported LCN is based on an intermediate thermal step, where the flux of a control light beam is converted into a temperature variation (hence triggering the phase transition) by the presence of a push-pull azo-dye in the LCN mixture [43,44].

To demonstrate multibeam polarization management, we exploit the different diffraction orders of a one-dimensional grating. We focused our attention on the transmitted (i.e., zero-order diffracted) and first-order diffracted beams, as represented in Fig. 1a. A polarimetric measurement has been performed by illuminating the sample with a linearly polarized He-Ne laser beam, focused to a spot size roughly equal to the grating size. The transmitted and diffracted beams were analyzed with a rotating linear polarizer and a power meter. Assuming that the grating does not degrade the degree of polarization of the incident light, this procedure allows to extract the transmitted/diffracted beam intensity and polarization state modulo the handedness; in other words, one can access the Stokes parameters with an indetermination over the sign of $S_3$. The results are plotted in Fig. 2a-d as points of different colors and shapes. Three different samples, characterized by different thicknesses of the grating stripes (250, 500, and 750 nm), have been investigated. The periodicity and the stripe width are kept fixed at 1.5 µm and 500 nm, respectively. Experimental data are plotted superimposed to a set of curves representing the result of numerical simulations performed by means of the rigorous coupled-wave analysis method (RCWA), following the formalism reported in previous works [45,46]. The code is available as a MATLAB package[i]. We assumed for the LCN ordinary and extraordinary refractive indices $n_{o,0} = 1.47$ and $n_{e,0} = 1.59$, respectively. The experimental data are consistent with the simulated ones, with an exception on the absolute level of the Stokes parameters of the diffracted beam, where a rescaling between the theory and the experiment is observed. This effect can be attributed to an uncertainty in the knowledge of the LCN refractive index distribution: indeed, slight misalignments of the molecules in the vicinity of the stripe boundary may have occurred. However, the trends of the experimental data as a function of the grating thickness follow very well the behavior expected from the calculation, for both horizontal and vertical incident polarized light.

These results indicate that control over the polarization state of diffracted beams has been attained, implementing a non-trivial polarization transformer from linear to elliptical, tilted polarization state. To visualize more clearly this effect, we reported in Fig. 2e the trajectory of the diffracted beam polarization state over the Poincaré sphere as the stripe thickness is increased, for the case of horizontally polarized incident light, i.e. for the case of Fig. 2b. It can be noticed that the polarization state spans the sphere in several regions, getting close to all the relevant points exception made for the V state. Notice that the range of grating thicknesses has been enlarged with respect to Fig. 2b, to better illustrate the potentials of the phenomenon.

As anticipated, the unicity of LCN anisotropic gratings is that they can be remotely tuned, implementing a full-optically controlled polarization switch acting on multiple channels. The setup employed for this study is sketched in Fig. 3a. Under vertically polarized incident light the diffracted beam is almost linearly polarized with the axis tilted at -45°. A linear polarizer is set orthogonally to the diffracted light polarization, and the resulting filtered light is detected and temporally analyzed by a fast camera. Remote control over the grating state is achieved through a light beam from a green laser diode illuminating the whole sample, chopped at a frequency of 100 Hz with a 50% duty cycle. Figure 3b reports the time trace of the diffracted beam intensity after passing through the polarizer, as a function of time. An observable effect can be noticed for driving powers as low as 1 mW; increasing the power level the effect becomes more intense while maintaining the overall time dependence. The effect occurs on two distinct time scales (Fig. 3c): a slow one, of the order of the chopping period (5 ms), and a fast one, of the order of a 0.2 millisecond or less (responses faster than 0.2 ms cannot be resolved due to the intrinsic transition time of the mechanical chopper).

In order to get insights on the nature of the switching mechanism that occurs in our device, we performed fully vectorial electromagnetic modeling and thermal dynamics modeling. First we consider electromagnetic modeling, where we parametrized the switching process with a parameter $x$ that represents the degree of molecular order of the LCN. To grasp the essence of the switching operation, the ordinary and extraordinary refractive indices are modeled by a simple linear relation:

$$n_o(x) = n_{o,0} + (\bar{n} - n_{o,0}) x$$

$$n_e(x) = n_{e,0} + (\bar{n} - n_{e,0}) x \qquad (1)$$

where $n_{o,0} = 1.47$, $n_{e,0} = 1.59$, and $\bar{n} = 1.51$. Here, $x = 0$ identifies the nematic LCN state, while $x = 1$ describes the perfectly isotropic state. In addition, due to the presence of a glass substrate, the LCN stripes undergo internal stresses that induce a possible trapezoidal-like shape modification of the stripe, as shown in Fig. 3a. To further simplify the model, we have however introduced only a modification of the stripe thickness, which can be regarded as an equivalent geometric effect:

$$t(x) = t_0 (1 + 0.05 \, x) . \qquad (2)$$

Here, the factor 0.05 is a reasonable estimate for the contraction ratio of thin LCN samples lying on a rigid substrate. In order to better understand the nature of the switching process, we have considered three different but complementary models: (i), a full model, i.e. a model described by equations (1) and (2) where x is varied from 0 to 1 ; (ii), a model where $t$ does not depend on $x$, i.e. $t = t_0$ ; (iii) a model where $n_o$ and $n_e$ do not depend on $x$, i.e. $n_o = n_{o,0}$ and $n_e = n_{e,0}$. Fig. 3d shows the polarization state of the diffracted channel as red-hue dotted traces on the Poincaré sphere, according to the models explained above. The size of the red-hue dots is proportional to the intensity of the beam, while the black-surrounded dot indicates the polarization state of the diffracted beam when $x = 0$ in both equations (1) and (2) (unperturbed stripe dimension and anisotropic refractive index). The polarization state produced by the models (i) and (ii) sweeps from a polarization state lying close to the -45° state to the V state expected when a grating constituted of an isotropic material is illuminated at normal incidence with V polarized light. Moreover, in case (i), the variation of the polarization state is also accompanied by a decrease of the intensity of the diffracted beam. When instead one assumes that the switching mechanism has contribution from the sole shape change effect, a completely different polarization modulation is expected: this is represented by path (iii), which spans the sphere along another direction, without a noticeable amplitude modulation effect.

It is interesting to observe that these two simple control mechanisms (induced by refractive index or by stripe thickness variation) lead to different effects, yet compatible with the experimental observations of signal modulation for the polarization-filtered diffracted beam. Consider for instance the switch-on time slot (Fig. 3b), where two opposite trends are present: a sudden signal decrease is followed by a slow increase. In the switch-off time slot the opposite occurs; this peculiar behavior can be interpreted in the spirit of the results of models (i)-(iii) above. In a first step, the polarization state, which at the origin is not perfectly linear, becomes close to a perfect linear state (first steps of process (iii)). Hence it is filtered more efficiently by the polarizer, and the detected intensity decreases. In a second step, the polarization state moves instead towards the V state (process (i) or (ii)), being thus able to pass through the polarizer. The increase of the detected signal registered in such a second phase is hence safely compatible with process (ii), and may be compatible with process (i) provided that the diffracted beam intensity decrease does not compensate for the polarization state change. It should be noticed that here we considered the full range $x = 0 ... 1$ for illustrative purposes; in a practical device, in the high-temperature induced paranematic state the LCN is likely not fully isotropic, and a restricted range for the $x$ values could be more relevant when comparing the model with the experiments.

Simultaneously with polarization modulation of the diffracted beam, also the transmitted beam experiences a polarization modulation. This is represented in Fig. 3d by the blue-hue dots (trace (iv)): starting from an elliptical polarization state (different from the polarization state of the diffracted beam), the transmitted beam reaches a linear polarization state when $x$ is swept from 0 to 1. This indicates that independent control over the polarization and amplitude of different channels – transmitted and diffracted channels – is achieved.

A further confirmation of the reliability of the actuation model is provided by a finite-element thermal dynamics analysis. Detailed information is reported in the Supplementary Material; in essence, the transient temperature distribution of the grating dictated by light absorption and heat diffusion is simulated, for both phases of switch-on and switch-off of the control beam. The thermalization dynamics, similar for both phases, shows a main thermalization time of 0.4 ms; within this time interval the temperature sweeps half of the span between the two steady-state values (room temperature for the "off" phase, and ~140 °C for the "on" phase). However, the process is not described by a single exponential; full thermalization occurs on the time scale of several milliseconds. This result supports the picture given above for the actuation mechanism: the fast feature observed in Fig. 3 matches the first stage of the thermal dynamics, while the slow one can be connected to the second stage of the dynamics.

In summary, we demonstrated that an anisotropic diffraction grating fabricated from a liquid crystal network through a nanopolymerization technique acts as a multichannel polarization handling device, which is remotely tunable by means of a control light beam. Although the polarization rotation performance still needs to be optimized, the device has several strength points: advanced intensity and polarization state targeting, submillisecond time response, absence of liquid components, and the possibility to address the working point in a wireless fashion. It might be envisaged that phase modulation is also possible by means of the present technique. These features place our device within the area of interest for several applications, such as polarimetry (to map the Stokes parameters into different spatial channels), augmented reality (where diffraction gratings are employed to superimpose the artificial image to the real one), and optical communications (to multiplex/demultiplex polarization-encoded channels).

**Supplementary Material**

Supplementary Material containing details of the sample fabrication process, additional data on the optical response, and thermal diffusion simulations, are available.

**Acknowledgements**

The authors acknowledge funding from the European Commission (EU-H2020 GA 654148 "Laserlab-Europe") and Ente Cassa di Risparmio di Firenze (grant 2015/0781).

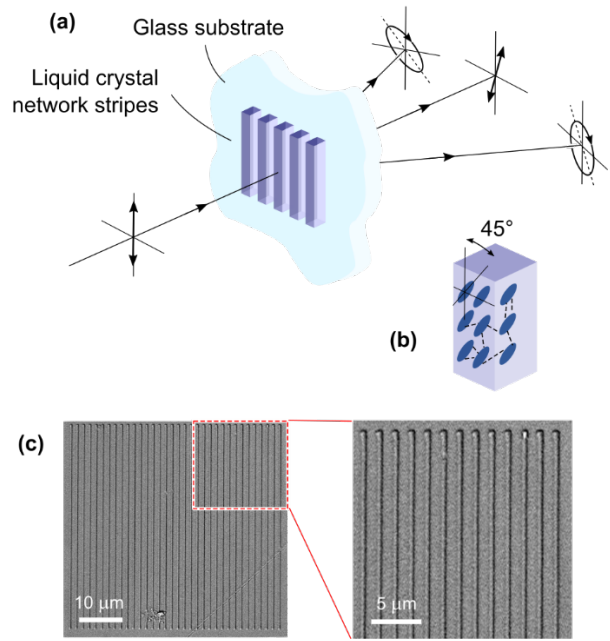

Figure 1. (a) Schematic of the device and illustration of the operation concept. The liquid crystal network (LCN) grating diffracts light; thanks to its anisotropy (b), it converts the linear polarization state of an orthogonally incident beam into general elliptic polarization states. Notably, the diffracted beam polarization differs from that of the zero-order transmitted beam, and can be targeted by appropriate shaping of the LCN stripes. The ellipsoids in (b) depict the LCN monomers, and the dashed lines indicate the molecular crosslinks. (c) SEM micrographs of a fabricated sample.

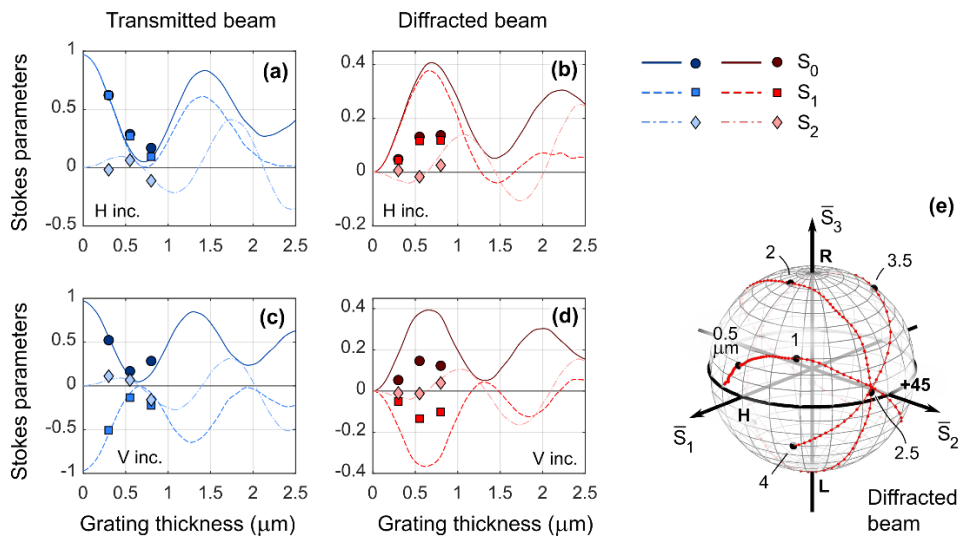

Fig. 2. (a)-(d) Polarization state of transmitted and diffracted beams, upon different incident polarization conditions (V and H inc. pol.), as a function of the grating stripe thickness. The traces represent the result of the simulation, while the dots represent experimental results. $S_{0-2}$ are the first three Stokes parameters. (e) Representation on the Poincaré sphere of the diffracted beam polarization state, assuming H incidence. The data are an extension of the simulated traces of (b); numbered labels identify the grating stripe thickness. $\bar{S}_{1-3}$ are the normalized Stokes parameters.

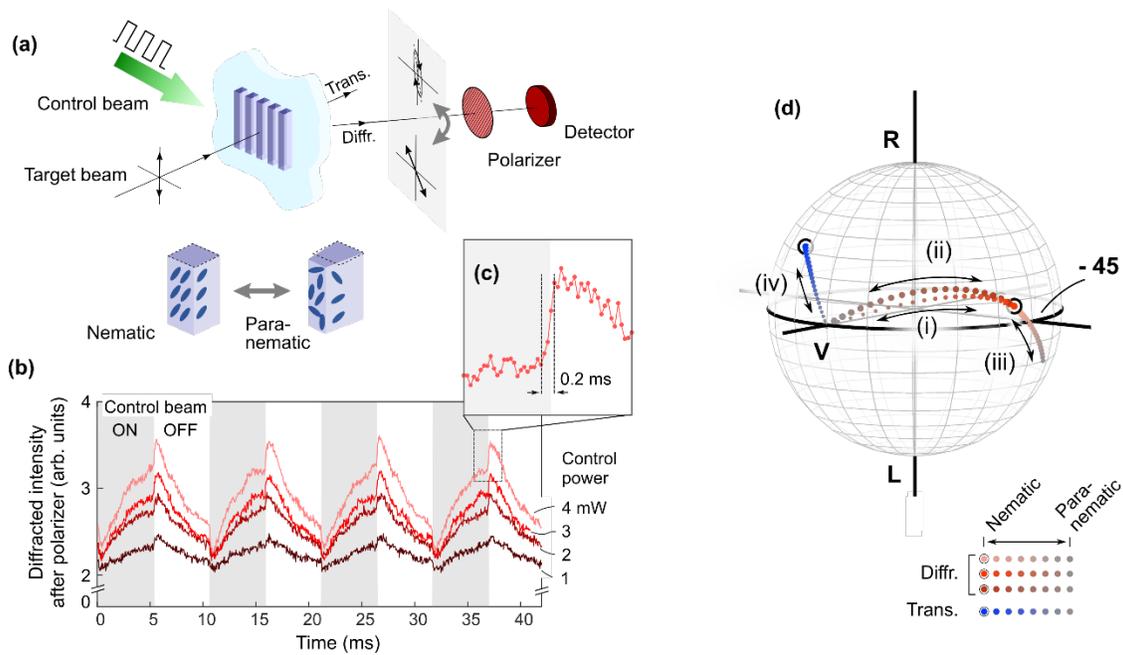

Fig. 3. (a) Schematic of the polarization modulation effect and of the setup employed for its measurement. Modulation originates from refractive index, optical anisotropy and shape change occurring across the nematic-paranematic phase transition. The grating has stripe width and thickness of 450 nm and 1200 nm, respectively. (b-c) Time trace of the polarization filtered diffracted signal as recorded by the detector. Gray regions highlight the time intervals where the control light is actuating the grating. (d) Representation on the Poincaré sphere of the polarization state evolution during the switching process. Paths (i)-(iii) illustrate the phenomenon on the diffracted beam, and have been calculated assuming different contributions to the switching mechanism (see text). Path (iv) illustrates the phenomenon on the transmitted beam. The size of the dots represents the intensity of the diffracted/transmitted beam.

---

[i] https://it.mathworks.com/matlabcentral/fileexchange/55401-ppml-periodically-patterned-multi-layer

*Supplementary Material for the article*

## Multichannel remote polarization control enabled by nanostructured Liquid Crystalline Networks

Simone Zanotto[1], Fabrizio Sgrignuoli[2,3], Sara Nocentini[2,4,5], Daniele Martella[4,6,7], Camilla Parmeggiani[4,5,6,7], and Diederik S. Wiersma[2,4,5,6]

1. Istituto di Nanoscienze - CNR, Laboratorio NEST, Piazza San Silvestro 12, 56127 Pisa, Italy
2. Dipartimento di Fisica ed Astronomia, Università degli Studi di Firenze, Via Sansone 1, 50019 Sesto Fiorentino (FI), Italy
3. Department of Electrical and Computer Engineering, Boston University, 8 Saint Mary's Street, Boston, MA, 02215-1300, United States of America
4. LENS - European Laboratory for Nonlinear Spectroscopy, Università degli Studi di Firenze, Via Nello Carrara 1, 50019 Sesto Fiorentino (FI), Italy
5. Istituto Nazionale di Ricerca Metrologica, Strada delle Cacce 91, 10135 Torino, Italy
6. Istituto Nazionale di Ottica – CNR, Via Nello Carrara 1, 50019 Sesto Fiorentino (FI), Italy
7. Dipartimento di Chimica "Ugo Schiff", Università degli Studi di Firenze, Via della Lastruccia 3-13, 50019 Sesto Fiorentino (FI), Italy

### 1. Details of the Liquid Crystalline Network fabrication

The Liquid Crystalline Network (LCN) grating is fabricated starting from a mixture of monomers whose molecular structures are depicted in Fig. S1. Mesogen, crosslinker and photoinitiator are commercial molecules while the azo-dye has been chemically engineered to be compatible with the two-photon absorption polymerization[1]. The nanofabrication potentiality of this mixture has been already systematically tested and explored in photonics as well in microrobotics. The molar ratio between the components is the following: mesogen 68%, crosslinker 30%, azo-dye 1%, photoinitiator 1%; this unpolymerized mixture has a nematic-isotropic phase transition occurring at 65 °C. The monomer mixture is infiltrated in a glass cell formed by two coverslips separated by 20 μm diameter spacing spheres. The inner surfaces of the cell have been spin-coated with a sacrificial polyimide layer and subsequently rubbed along the appropriate direction to impress the LC alignment. The cell has been subsequently mounted on the stage of a commercial DLW system (Nanoscribe GmbH) where the nanopatterning process took place. After the writing step, the cell has been opened and immersed in a hot (50°C) isopropanol bath for 20 minutes in order to remove the unpolymerized mixture.

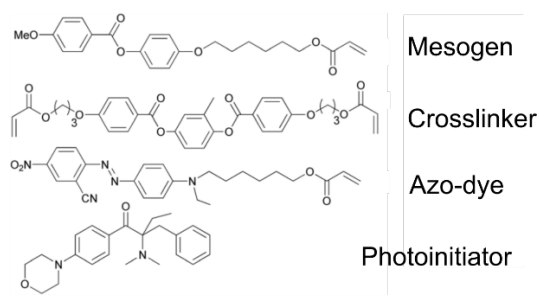

Fig. S1. Structure of the molecules employed for the fabrication of the LCN grating.

## 2. Grating stripe local morphology

The local morphology of the grating stripe, which eventually impacts the grating optical quality, can be observed in Fig. S2. The written lines do not show any roughness at length scales comparable to the red light at which the grating has been characterized. This observation is consistent with the good quality of the diffraction pattern observed in the far-field. The diagonal line is a spurious effect due to the rubbing process involved in the cell construction; here it can also be regarded as an indication of the liquid crystal molecule alignment with respect to the grating stripe orientation.

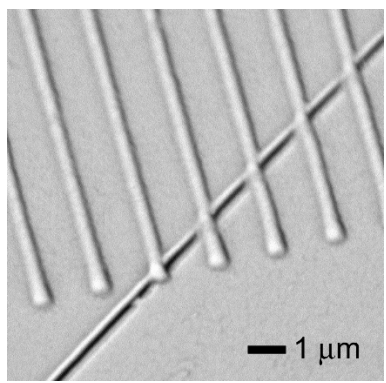

Fig. S2. High resolution scanning electron microscope image of the grating stripes.

## 3. Thermal dynamics of the grating upon light illumination

Thermal dynamics analysis of the grating upon control light illumination has been performed by means of finite-element modeling, through the *Heat Transfer* module of the FEM software COMSOL. A 3D domain containing the whole grating is simulated, in both warm-up and cool-down phases. To simulate the light absorption and heating of the grating, a heat source is placed in the grating region. By setting the source power to a level compatible with the actual experiment conditions, a steady state temperature of 140 °C is reached, a value close to the nematic-paranematic phase transition[2]. The LCN material is simulated as a medium with an anisotropic, temperature-dependent thermal conductivity: $k_{par}(T) = k_{par,0}$ $\forall$ T; $k_{perp}(T) = k_{perp,0}$ for T < 50 °C, $k_{perp}(T) = k_{par,0}$ for T > 180 °C, $k_{perp}(T) = k_{perp,0} + (k_{par,0} - k_{perp,0}) \cdot (T - 50)/130$ for 50°C < T < 180°C; where $k_{perp,0}$ = 0.3 W m$^{-1}$ K$^{-1}$ and $k_{par,0}$ = 3 W m$^{-1}$ K$^{-1}$. Here, $k_{par}(T)$ and $k_{perp}(T)$ are the thermal conductivity tensor components, respectively, parallel and perpendicuar to the LC director. The time dependence (Fig. S3) shows that half of the temperature span is covered in a short time, 0.4 ms; to reach full thermalization, a much longer time is needed instead (the time evolution is not a single exponential). This result supports the picture given in the main text for the actuation mechanism: the fast feature observed in Fig. 3 of the main text matches the first stage of the thermal dynamics, while the slow one can be connected to the second stage of the thermalization dynamics.

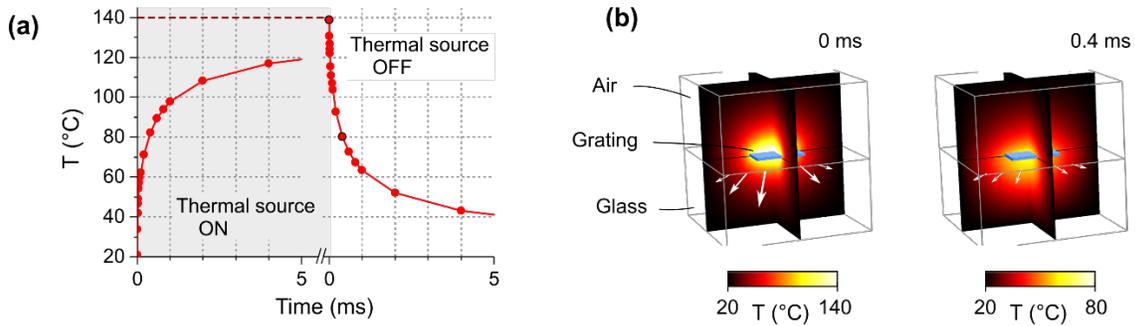

Fig. S3. (a) Time dynamics determined by finite element calculations. The average temperature of the grating spans half of the temperature excursion in about 0.4 ms. Steady-state temperatures are 140 °C and 20 °C when the grating is excited or not excited, respectively. (b) The heat flow (white arrows) and the temperature (color scale) are plotted at two representative times during the cool-down process. Arrows have the same scale in the two images.

## 4. Response of the grating to other input polarization states

To complement the data provided in the main text, we illustrate in Fig. S4 the polarization state of the diffracted beam when the input beam is linearly polarized along the vertical direction (V), or when it is circularly polarized with either left (L) or right (R) handedness. The quantities $\bar{S}_{1,2,3}$ identify the normalized Stokes parameters. It can be noticed that the path on the Poincaré sphere is almost identical to that observed in Fig. 2e of the main text, except for rigid rotations.

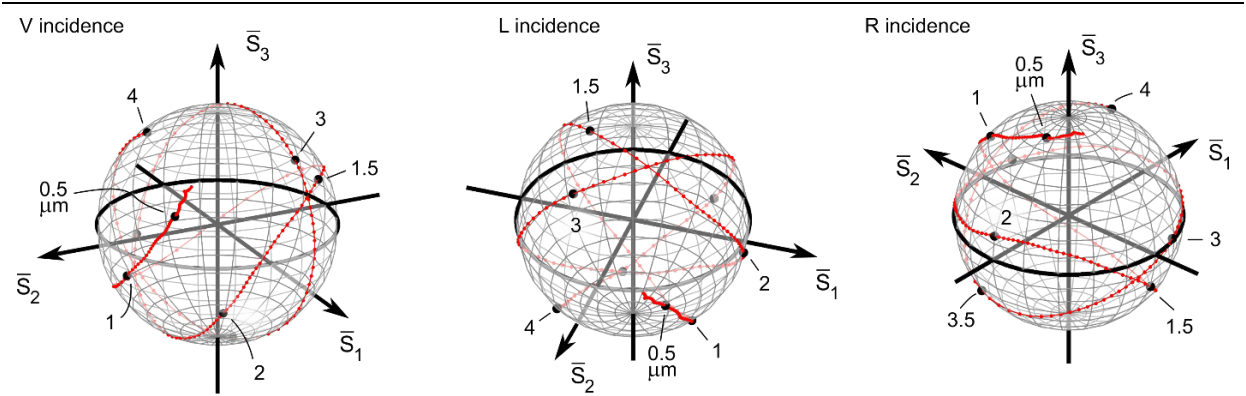

Fig. S4. Polarization state of the diffracted beam upon vertical linearly polarized light (V), and circularly polarized light of left (L) and right (R) handedness. The numbers indicate the thickness of the grating stripes (see Fig. 2 of the main text and the related discussion).

Finally we provide the dynamic response of the tunable grating for different input polarizations. In Fig. S5 we report the generalization of the data of Fig. 3d of the main text: the blue dotted trace identifies the state of polarization of the transmitted beam, while the red dotted trace identifies the state of polarization of the diffracted beam. As in the main text, the size of the trace dots indicate the intensity of the beam. The arrow illustrate the switching process, starting from the nematic state of the LCN (control off) to the paranematic state (control on). Only the complete model (shape change plus refractive index change, see the main text for details) has been considered here. Notice that the diffracted beam does not reach perfect circular polarization even when the grating is in the fully "on" state.

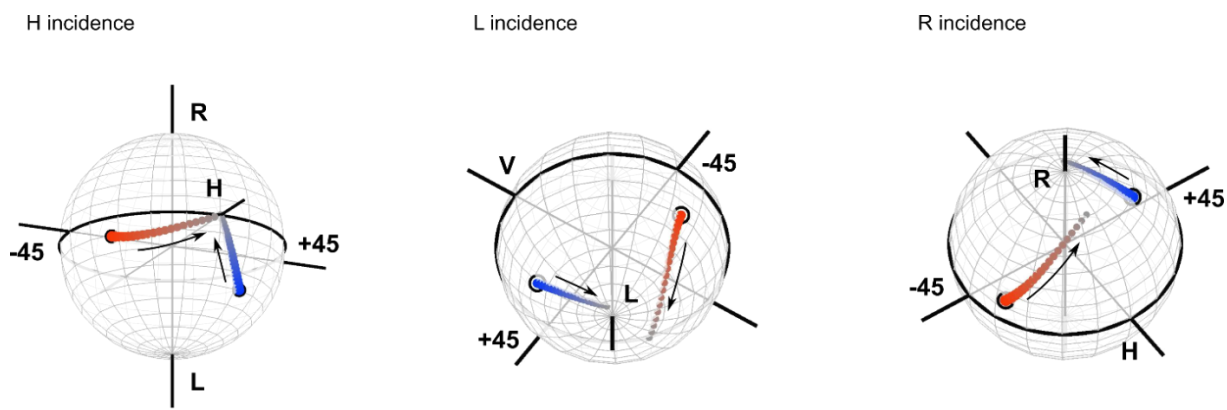

Fig. S5. Control over the polarization state of the transmitted (blue) and diffracted (red) beams upon LCN actuation, for different polarization states of the incident beam.